\documentclass{amsart}
\usepackage{amssymb,amsthm,euscript,amsmath,graphicx}
\usepackage[english]{babel}
\usepackage{t1enc}

\newcommand{\be}{\begin{equation}}
\newcommand{\ee}{\end{equation}}
\newcommand{\ben}{\begin{equation*}}
\newcommand{\een}{\end{equation*}}
\newcommand{\bg}{\begin{gather}}
\newcommand{\eg}{\end{gather}}
\newcommand{\bea}{\begin{eqnarray}}
\newcommand{\eea}{\end{eqnarray}}
\newcommand{\eean}{\end{eqnarray*}}
\newcommand{\bean}{\begin{eqnarray*}}
\newcommand\re[1]{(\ref{#1})}
\newcommand\ol[1]{\overline{#1}}

\newcommand\Ga{\Gamma}
\newcommand\ep{\epsilon}

\begin{document}

\title{Generic stability of dissipative non-relativistic and relativistic fluids}
\author{ P\'e\-ter V\'an}
\address{Department of Theoretical Physics\\KFKI, Research Institute of Particle and Nuclear Physics, Budapest, Hungary\\ and
Department of Energy Engineering\\Budapest University of Technology and Economics
}
\email{vpethm@rmki.kfki.hu}

\date{\today}
\begin{abstract}
The linear stability of the homogeneous equilibrium of non-relati\-vis\-tic fluids with mass flux and special relativistic fluids with the absolute value of the energy vector as internal energy is investigated.
It is proved that the equilibrium is asymptotically stable in both cases due to purely thermodynamic restrictions; the only requirements are the thermodynamic stability and the nonnegativity of the transport coefficients.  
\end{abstract}
\maketitle


\section{Introduction}

Thermodynamics is a theory of stability. The existence of the entropy function, a concave thermodynamic potential  that increases in insulated systems is a stability requirement. Similarly, the existence of thermodynamic potentials with particular convexity properties and their tendency to a definite extremum in case of different homogeneous boundary conditions are stability requirements. This is a rather old, but essential point of view regarding the nonequilibrium thermodynamics of macroscopic systems, both continuous and homogeneous. The similarity of the mathematical structure of the continuum thermodynamic theories and the conditions in Lyapunov's direct method (see e.g. \cite{RouAta77b}), was pointed out and investigated by several authors \cite{ColMiz67a,GlaPri71b,Gur75a,Daf79a}.   Moreover, a  complete and rigorous dynamic reinterpretation of ordinary equilibrium thermodynamics, the thermodynamics of homogeneous bodies  from the point of view of stability is given in \cite{Mat05b} (see also the references therein). According to this point of view the homogeneous equilibrium of macroscopic continua, the thermodynamic  equilibrium in case of  homogeneous boundary conditions is asymptotically stable. Pure thermodynamic requirements  are  the only conditions  without any further ado. In the following we call this essential property as \textit{generic stability}.

Let us give some clarifying remarks to the statement above. First of all generic stability can be violated in everyday situations. The loss of thermodynamic stability, that is the violation of concavity, the existence of phase boundaries, indicate that changes of the external conditions can trigger structural changes in the material. As a  second remark let us emphasize, that generic  stability is related to specific equilibria and not to steady states.  The stability or instability of steady states is not related directly to generic stability \cite{GlaPri71b}. Finally we emphasize that the existence of a Lyapunov function ensures that generic stability is nonlinear, and the  linear stability of the same equilibria is  only a necessary condition. 

This stability point of view of thermodynamics gives a sound  and simple interpretation of the Second Law and shows clearly why the Second Law is fundamental for any theory of physics. Without this basic stability property of materials one could not measure anything, repeated experiments could give rather different results, small changes of the external conditions could drive physical  systems far from their initial state.  

The problem of generic stability is well known and crucial for dissipative relativistic fluids. The simplest relativistic generalization of the Fourier-Navier-Stokes system of equations, the theory of Eckart \cite{Eck40a3} is proved to be unstable and acausal and therefore not viable \cite{HisLin85a}. The same has been proved for the so-called first order theories in general, where entropy four vector depends on the classical basic state variables \cite{His86a}. 

The    other conceptual issue of relativistic fluids, the requirement of causality, initiated a whole new family of dissipative relativistic fluids, where the basic state space is extended by different ways. In several cases the extension involves the dissipative part of the flux of energy-momentum as new variables of the theory. When the  extension results in  symmetric hyperbolic evolution equations, then causality is ensured.
The price of the extension is the increased complexity of the equations and the larger number of necessary material parameters. There are a number of such kind of theories in the literature \cite{Mul69a,Isr76a,IsrSte79a,LiuAta86a,PavAta82a,Ott05b}. Only a few of them was investigated from the point of view of symmetric hyperbolicity and generic stability.

Regarding the connection of causality and  hyperbolicity there are arguments that from a physical point of view parabolic theories can be causal as long as  they are  stable \cite{Fic92a,KosLiu00a,Cim04a,VanBir08a}, because in that case the propagation speed of the validity range of the theory can be smaller than the speed of light. On the other hand, stability and symmetric hyperbolicity are not necessarily connected, a stable parabolic theory can be extended.

The example of the Israel-Stewart theory, the best known extended theory in the literature, demonstrates the rather involved relations of causality, stability and hyperbolicity of relativistic fluids. The original non-perturbed equations in these theories are not known to be symmetric (let alone causal) for arbitrary fluid states \cite{GerLin91a}. However, conditions of the symmetric hyperbolicity of the perturbation equations  are equivalent to the conditions of linear stability of the homogeneous equilibrium. What is really disturbing here are these  stability and hyperbolicity conditions themselves  \cite{HisLin83a,HisLin88a} (See Appendix A). These conditions give restrictions on the equations of state of the fluid beyond normal thermodynamic requirements, conditions    for the newly introduced additional material parameters (e.g. they cannot be constant) and also restrictions regarding their relations. Moreover, the physical content of these difficult relations is completely obscure.
As the Israel-Stewart theory is the most popular in recent applications to the fluid mechanic description of heavy-ion collisions all these conceptual problems are can have an experimental feedback 
\cite{Mur04a,HeiAta06a,BaiRom07a,DenAta08a,DenAta08a1,HeiAta06a,HeiSon08a,MolHou08a,Mol08m}. 

Recently we have suggested a new approach to the problem. Our suggestion is based on a novel concept of internal energy \cite{VanBir08a}. The form of the internal energy was derived from a detailed investigation of the compatibility of the basic balances and the entropy inequality by modern thermodynamic methods \cite{Van08a}. We have investigated also the generic stability of the fluid equations in the Eckart frame, and proved that thermodynamic stability and the positivity of the usual transport coefficients (heat conduction, viscosity) will ensure the stability without any further ado \cite{VanBir08a}. Moreover, we have given a simple extension of our internal energy that gives relaxation equations for the dissipative fluxes similarly to the Israel-Stewart theory \cite{BirAta08a}.

In this paper we investigate necessary conditions of the asymptotic stability of homogeneous equilibrium of insulated simple one component fluids. First we will show the linear stability of  a generalized form of one component nonrelativistic  fluids with a diffusion flux. These fluids  were recently promoted by Brenner \cite{Bre05a,Bre05a1} and their thermodynamic suitability  was shown in the frame of GENERIC \cite{Ott05b} and Classical  Irreversible Thermodynamics \cite{BedAta06a}. The existence of diffusion flux in nonrelativistic fluids is equivalent to a general flow frame in relativistic fluids, when the velocity field of the continuum is not necessarily fixed to a conserved quantity. Then we investigate the generic stability of our relativistic dissipative fluid theory. We show that the conditions of the linear stability in a general frame are the thermodynamic stability and the positivity of the transport coefficients (heat conduction, viscosities and diffusion) without any further ado. Then we summarize and discuss our results.

\section{Non-relativistic fluids}

Here we develop a generalized Fourier-Navier-Stokes theory, from the point of view of non-equilibrium thermodynamics. Because we want to give a parallel treatment with the special relativistic case therefore we give the basic balances with the help of densities, introduce a conductive flux term in the balance of mass-particle number and apply a notation with indices in both cases. In case of non-relativistic fluids the  indices are $i,j,k,l,...$ and are running from $1,..,3$. The dot denotes the substantial time derivative $\dot a = \partial_t a +v^i\partial_i a$, where $\partial_t$ is the partial time derivative.
\subsection{Basic equations}

The balance of particle number will be given instead of the balance of mass
\begin{equation}
\dot n +n\partial_iv^i+\partial_ij^i=0.
\label{mbal}\end{equation}

In our one component fluid the particle number density $n$ is related to the density as  $\rho=mn$, where $m$ is  the molar mass. The balance of total energy is
\begin{equation}
\dot e + e\partial_iv^i+\partial_i l^i=0.
\label{ntebal}\end{equation}

Here $e$ is the conserved total energy density, $l^i$ is the energy flux. The balance of the momentum, the Cauchy equation, will be written as
\begin{equation}
\dot p^i + p^i\partial_jv^j+\partial_j\left(P^{ij}+j^jmv^i\right)=0^i.
\label{nmombal}\end{equation}

Here $p^i$ denotes the momentum density, we have assumed a continuum without internal moment of momentum, therefore  $P^{ij}$, the pressure tensor is symmetric. There is an extra contribution to the momentum flux due to the diffusion flux. We assume that the momentum is related to the velocity and the mass as $p^i=\rho v^i$. The internal energy density is the difference of the total and the kinetic energy densities as $\ep=e-\frac{\rho}{2}v^2$, therefore  the balance of internal energy is calculated from \re{mbal}-\re{nmombal} as 
\begin{equation}
 \dot \ep + \ep\partial_iv^i+\partial_i\left(l^i-P^{ji}v_j-j^i\frac {mv^2}{2}\right)=
 -P^{ij}\partial_jv_i.
\label{niebal}\end{equation}

Here we can recognize the flux of the internal energy, the heat flow as
\begin{equation}
q^i=l^i-P^{ji}v_j-j^i\frac {mv^2}{2}.
\end{equation}

We assume local equilibrium, therefore the entropy is the function of the internal energy and the density. The Gibbs relation identifies the derivatives of the entropy and the potential relation defines the static pressure in our density based treatment (see e.g. \cite{Mat05b}).
\begin{equation}
 d\ep = Tds+\mu dn, \qquad \text{and }\qquad
 \ep=Ts+\mu n -p.
\end{equation}

In the following we exclude phase transitions and assume thermodynamic stability, a concave entropy function. That can be ensured by the following inequalities
\begin{gather}
\frac{\partial}{\partial \ep}\frac{1}{T} <0,\quad
\frac{\partial}{\partial n}\frac{\mu}{T} >0, \nonumber\\
\frac{\partial}{\partial \ep}\frac{1}{T}\frac{\partial}{\partial n}\frac{\mu}{T}
-\frac{\partial}{\partial n}\frac{1}{T}\frac{\partial}{\partial \ep}\frac{\mu}{T}=
\frac{\partial}{\partial \ep}\frac{1}{T}\frac{\partial}{\partial n}\frac{\mu}{T}
+\left(\frac{\partial}{\partial n}\frac{1}{T}\right)^2\leq 0.
\label{nrtdstab}\end{gather}

Here we exploited the equality of the mixed partial derivatives of the entropy $ \frac{\partial}{\partial n}\frac{1}{T}=-\frac{\partial}{\partial \ep}\frac{\mu}{T}$.
Let us assume that the entropy flux $J^i$ has the classical form \cite{GroMaz62b} \begin{equation}
J^i = \frac{q^i}{T} - j^i \frac{\mu}{T}.
\end{equation}

Let us remark that this form of the entropy flux is the consequence of the local equilibrium and first order nonlocality and can be derived from the Second Law \cite{Van02a}.

Therefore the entropy balance can be written as:
\begin{multline}
\dot s(\ep,n) + s \partial_i v^i + \partial_i J^i = ... \\
         =-j^i \partial_i \frac{\mu}{T}+
        q^i\partial_i \frac{1}{T}-
        \frac{1}{T}\left(P^{ij}-p\delta^{ij}\right)\partial_iv_j\geq 0.
\label{nrSL}\end{multline}

The usual thermodynamic fluxes are the heat flux $q^i$ and the diffusion flux $j^i$ and the dissipative pressure $P^{ij}-p\delta^{ij}$. The corresponding thermodynamic forces are the gradients of the entropic intensives $\partial_i \frac{1}{T}$ and $-\partial_i \frac{\mu}{T}$, and the velocity gradient $\partial_iv_j$ respectively. Usually we assume a linear relationship between the thermodynamic fluxes and forces. For isotropic materials the representation theorems of isotropic functions give scalar coefficients in the Fourier-law, in the Fick-law and in the Newtonian pressure as
\begin{eqnarray}
 q^i &=& \lambda \partial_i \frac{1}{T}- \chi \partial_i \frac{\mu}{T}, \label{nfour} \\
 j^i &=& \chi \partial_i \frac{1}{T}- \xi \partial_i \frac{\mu}{T},\label{nfick}\\
 P^{ij}-p\delta^{ij}=: \Pi^{ij} &=& -\eta(\partial^iv^j + \partial^jv^i- \frac{2}{3} \partial_kv^k\delta^{ij}) - \eta_v \partial_kv^k\delta^{ij}. \label{nnewt} \end{eqnarray}

Here $\Pi^{ij}$ denotes the dissipative part of the pressure. $\hat\lambda=\lambda T^{-2}$ is the Fourier heat conduction coefficient, $\xi$ is the diffusion coefficient and $\chi$ is the coupling coefficient of the Soret-Dufour effect, the coupling between heat conduction and diffusion. The cross coefficients are equal according to the reciprocity relations of Onsager. $\eta$ is the shear viscosity and $\eta_v$ is the bulk viscosity, respectively. The inequality of the Second Law \re{nrSL}, that is the nonnegativity of the entropy production restricts the conductivity coefficients, therefore
\begin{equation}
\lambda>0, \qquad
\xi>0,  \qquad
\lambda\xi-\chi^2>0,  \qquad
\eta>0,  \qquad \eta_v>0.
\label{poscoeff}\end{equation}

In the following investigations we will neglect the Soret-Dufour effect and assume that $\chi=0$.

\subsection{Equilibrium}

In thermodynamic (generic) equilibrium the substantial time derivatives and the dissipative fluxes
$q^i$, $j^i$ and $P^{ij}-p\delta^{ij}$, are zero. Therefore from \re{nfour} and \re{nfick} it follows, that $T(\ep,n)=const.$ and $\mu(\ep,n)=const.$. The inequalities of thermodynamic stability \re{nrtdstab} ensures the local invertibility of these functions, therefore in generic equilibrium $\ep=const.$ and $n=const.$, too. From \re{nnewt} we get, that the velocity field is divergence and rotation free, that is $\partial_jv^j=0$ and $\partial_iv^j -\partial_jv^i=0$. In the following we  investigate the linear stability of the homogeneous equilibrium, where the relevant fields are constant in time and homogeneous. Moreover, as the velocity field in our equations is a relative velocity related to an inertial observer, we may assume without restricting the generality that the equilibrium velocity field is zero.  Therefore the equilibrium densities and fluxes are
\begin{gather}
n(x^j,t)=\ol n =const., \qquad
\ep(x^j,t) =\ol\ep=const., \qquad
v^i(x^j,t) = 0^i, \nonumber\\
j^j(x^i,t)=0^j, \quad
q^j(x^i,t)=0^{j}, \qquad
\Pi^{ij}(x^k,t)=P^{ij}(x^k,t)-p(\ol\ep,\ol n)\delta^{ij}=0^{ij}.
\label{nreq}\end{gather}

\subsection{Linearization}
Let us denote the perturbed fields by $(\delta n,\delta\ep, \delta v^i,\! \delta j^i,\! \delta q^i, \delta\Pi^{ij})$. The linearization of the balances \re{mbal}-\re{nmombal} around homogeneous equilibrium results in
 \bea
 0 &=& \dot{\delta n} +
        \ol n \partial_j \delta v^j+\partial_j\delta j^j=0, \label{lmbal}\\
 0 &=& \dot{\delta \ep} +
        (\ol\ep+\ol p) \partial_j \delta v^{j}+\partial_j \delta q^j, \label{lniebal}\\
 0^{i} &=& m\ol n\dot{\delta v^i}  +
                \partial_i  p +
                \partial_j \delta\Pi^{ij} =
        m\ol n+
               \left(\ol{\frac{\partial p}{\partial\ep}}  \partial^i\delta\ep+
                \ol{\frac{\partial p}{\partial n}} \partial^i\delta n \right) +
        \partial_j \delta\Pi^{ij},
        \label{lnmombal}\\
 0^{i} &=& \delta q^i -
        \lambda \partial^i \frac{1}{T}=
        \delta q^i -
        \lambda\left(\ol{\frac{\partial}{\partial\ep}\frac{1}{T}}  \partial^i\delta\ep+
                \ol{\frac{\partial}{\partial n} \frac{1}{T}}  \partial^i\delta n \right), \label{lnfour}\\
 0^{i} &=& \delta j^i +\xi\ \partial^i \frac{\mu}{T}=\delta j^i +\xi\ \left(\ol{\frac{\partial}{\partial\ep}\frac{\mu}{T}}  \partial^i\delta\ep+
                \ol{\frac{\partial}{\partial n} \frac{\mu}{T}}  \partial^i\delta n \right), \label{lnfick}\\
 0 &=&  \delta\Pi^{ij} +
        {\eta}_v \partial_k \delta v^k \delta^{ij} +
        \eta(\partial^i\delta v^j +
                \partial^j\delta v^i+
                \frac{2}{3}\partial_k \delta v^k \delta^{ij} ).
        \label{lnewtm}
\eea

The overline denotes the equilibrium value of the corresponding functions, e.g. $\ol p=p(\ol\ep,\ol n)$.

In order to identify possible instabilities we select out
exponential plane-wave solutions of the perturbation equations:
$\delta Q =  Q_0 e^{\Gamma t + i k x}$, where $Q_0$ is constant.
As our equilibrium background state is a fluid at rest, the substantial time derivative is the partial time derivative $d/dt  =\partial_t$.

With these assumptions the perturbation equations follow as
 \begin{eqnarray*}
 0 &=& \Ga{\delta n} + i k \ol n \delta v^x + i k \delta j^x,
    \label{nrlin1}\nonumber\\
 0 &=& \Ga{\delta \ep} +(\ol\ep+\ol p) i k \delta v^x + i k \delta q^x,
    \label{nrlin2}\nonumber\\
 0 &=& \Ga m\ol n\delta v^x +
        i k \left(\ol{\frac{\partial p}{\partial\ep}} \delta \ep +
            \ol{\frac{\partial p}{\partial n}}\delta n \right) +
        i k \delta\Pi^{xx},\label{nrlin3x}\nonumber\\
 0 &=& \Ga m\ol n\delta v^y +
        i k \delta\Pi^{xy},\label{nrlin3y}\nonumber\\
 0 &=& \Ga m\ol n\delta v^z +
        i k \delta\Pi^{xz},\label{nrlin3z}\nonumber\\
  \end{eqnarray*}
 
\begin{eqnarray} 
0 &=& \delta q^x -
        i k \lambda\left(\ol{\frac{\partial}{\partial\ep}\frac{1}{T}}\delta\ep+
             \ol{\frac{\partial}{\partial n} \frac{1}{T}}\delta n \right),
    \label{nrlin4x}\nonumber\\
 0 &=& \delta q^y = \delta q^z, \label{nrlin4yz}\nonumber\\
 0 &=& \delta j^x +
        i k \xi\left(\ol{\frac{\partial}{\partial\ep}\frac{\mu}{T}}\delta\ep+
            \ol{\frac{\partial}{\partial n}\frac{\mu}{T}}\delta n \right),
    \label{nrlin5x}\nonumber\\
 0 &=& \delta j^y = \delta j^z, \label{nrlin5yz}\nonumber\\
 0 &=&  \delta\Pi^{xx} + ik\left( \frac{8}{3}\eta+\eta_v\right)\delta v^x,
    \label{nrlin6xx}\nonumber\\
 0 &=&  \delta\Pi^{xy} + ik \eta\delta v^y,    \label{nrlin6xy}\nonumber\\
 0 &=&  \delta\Pi^{xz} + ik \eta\delta v^z,    \label{nrlin6xz}\nonumber\\
 0 &=&  \delta\Pi^{yy} + ik \eta_v\delta v^y,  \label{nrlin6yy}\nonumber\\
 0 &=&  \delta\Pi^{zz} + ik \eta_v\delta v^z,  \label{nrlin6zz}\nonumber\\
 0 &=&  \delta\Pi^{zy}. \label{nrlin6zy}
\eea

We can put the equations above into the following matrix form
 \be
  M^A_{\;\;B} \delta Q^B = 0.
\label{nrmeq}\ee

Here $\delta Q^B$ represents the list of 17 fields which describe the
perturbation of the Fourier-Fick-Navier-Stokes fluid:
\begin{gather*}
 \delta Q =
    (\delta n, \delta\ep, \delta v^x, \delta q^x, \delta\Pi^{xx}, \delta         j^x;\\
    \delta v^y,\delta \Pi^{xy},\delta \Pi^{yy}; \;\;
    \delta v^z, \delta \Pi^{xz},\delta \Pi^{zz};\\
    \delta \Pi^{yz}, \delta q^y, \delta q^z, \delta j^y, \delta j^z).
\end{gather*}

Then the 17x17 matrix ${\bf M}_{nr}$ can be written in the block diagonal
form
 \be
  {\bf M}_{nr} = \begin{pmatrix}
  {\bf N}_{nr} & 0       & 0       & 0       \\
  0       & {\bf R}_{nr} & 0       & 0       \\
  0       & 0       & {\bf R}_{nr} & 0       \\
  0       & 0       & 0       & \bf I \end{pmatrix},
 \label{Mnr}\ee

\noindent where {\bf I} is an 5x5 identity matrix and the submatrices ${\bf R}_{nr}$ and ${\bf N}_{nr}$ are defined as
follows
 \be
  {\bf R}_{nr} = \begin{pmatrix}
  mn\Ga       & ik    & 0\\
  ik\eta        & 1     & 0\\
  ik{\eta}_v    & 0     & 1\\
  \end{pmatrix},
 \label{Rnr}\ee

\be
  {\bf N}_{nr} = \begin{pmatrix}
  \Ga            & 0                     & ikmn      & 0    & 0 & ik \\
  0              & \Ga                   & ik(e+p)     & i k  & 0 & 0  \\
  ik{\frac{\partial p}{\partial n}} & ik{\frac{\partial p}{\partial\ep}}         & \Ga\rho & 0  & ik & 0 \\
  -ik\lambda{\frac{\partial}{\partial n}\frac{1}{T}}
        & -ik\lambda{\frac{\partial}{\partial\ep}\frac{1}{T}}
        & 0 & 1 & 0 & 0\\
  ik\xi{\frac{\partial}{\partial n}\frac{\mu}{T}}
        & ik\xi{\frac{\partial}{\partial\ep}\frac{\mu}{T}}
        & 0 & 0 & 0 & 1\\
  0     & 0 & ik\tilde{\eta}    & 0  & 1 & 0
  \end{pmatrix},
 \label{Nnrm}\ee

\noindent where $\tilde{\eta}= 8\eta/3 +\eta_v$ and we have removed the overline above the equilibrium quantities.

Exponentially growing plane-wave solutions of \re{meq} emerge
whenever $\Gamma $ and $k$ satisfy the dispersion relation
 \be
 \det {\bf M}_{nr} = (\det {\bf N}_{nr})(\det {\bf R}_{nr})^2=0
\ee with a positive real $\Gamma$. The roots of this equation are
the roots obtained by setting the determinants of either ${\bf N}_{nr}$ or
${\bf R}_{nr}$  to zero.

The determinant of ${\bf R}_{nr}$ gives the condition
 $$
\rho \Ga + {\eta} k^2 = 0,$$

\noindent which results in a negative $\Ga$.

The determinant of ${\bf N}_{nr}$ gives the following dispersion relation
 \begin{gather*}
 mn \Ga^3+\\
 k^2 \left(\eta -
        \lambda \rho \frac{\partial}{\partial \ep}\frac{1}{T} +
        \xi mn\frac{\partial}{\partial n}\frac{\mu}{T}\right)\Ga^2 +\\
 k^2 \left( k^{2}\left[\lambda\xi mn\Delta +
                \eta\left(\xi \frac{\partial}{\partial n}\frac{\mu}{T}
                -\lambda\frac{\partial}{\partial\ep}\frac{1}{T}\right)\right]+
          (\ep+p)\frac{\partial p}{\partial\ep} +
          mn \frac{\partial p}{\partial n}\right) \Ga  +\\
 k^6\lambda\xi\eta\Delta+
        k^4\xi(\ep+p)\left(\frac{\partial p}{\partial\ep}
                \frac{\partial}{\partial n}\frac{\mu}{T}-
                \frac{\partial p}{\partial n}
                \frac{\partial}{\partial\ep}\frac{\mu}{T}\right)+
        k^4mn\lambda\left(\frac{\partial p}{\partial\ep}
                \frac{\partial}{\partial n}\frac{1}{T}-
                \frac{\partial p}{\partial n}
                \frac{\partial}{\partial\ep}\frac{1}{T}\right)= 0.
 \end{gather*}

where
$$
\Delta=-\frac{\partial}{\partial\ep}\frac{1}{T}
                \frac{\partial}{\partial n}\frac{\mu}{T}+
                \frac{\partial}{\partial\ep}\frac{\mu}{T}
                \frac{\partial}{\partial n}\frac{1}{T}\geq 0,
$$

\noindent due to the concavity of the entropy.

According to the Routh-Hurwitz criteria \cite{KorKor00b}, the real
parts of the roots of a third order polynomial $a_0 x^3 +
a_1 x^2 + a_2 x + a_3=0$ are negative whenever \bea
 a_0 &>& 0, \nonumber\\
 a_1 &>& 0, \nonumber\\
 a_1 a_2 - a_0 a_3&>& 0.
\label{RHnr}\eea

We can see, that the first two conditions of \re{RH} are fulfilled, because the positivity of the density and the transport coefficients and the concavity of the entropy.
The third condition after some calculations gives the following expression
\begin{multline*}
a_1 a_2 - a_0 a_3 = \\
k^6\left(-\eta\lambda \frac{\partial}{\partial \ep} \frac{1}{T}
            \left[\eta+\lambda mn \frac{\partial}{\partial \ep}\frac{1}{T}\right]+\right.\\
        \xi\left[\eta^2\frac{\partial}{\partial n}\frac{\mu}{T} -
                2\eta\lambda mn\frac{\partial}{\partial n}\frac{\mu}{T}
                \frac{\partial}{\partial\ep}\frac{1}{T} -
                (mn)^2\lambda^2\frac{\partial}{\partial\ep}\frac{1}{T}\Delta                   \right]+\\
        \xi^2mn\frac{\partial}{\partial n}\frac{\mu}{T}\left.\left[
                \eta\frac{\partial}{\partial\ep}\frac{1}{T}+
                mn\lambda\Delta\right]\right)+\\
k^4\left(\xi mn\left[(\ep+p)\frac{\partial}{\partial n}\frac{1}{T}-
        \frac{\partial}{\partial n}\frac{\mu}{T}\right]^2 + \right.\\
  (\ep+p)^2\left[mn\lambda \left(\frac{\partial}{\partial\ep}\frac{1}{T}\right)^2-
        \eta \frac{\partial}{\partial\ep}\frac{1}{T}\right] +
         (\ep+p)2mn\frac{\partial}{\partial n}\frac{1}{T}\left[
                -\eta +mn\lambda\frac{\partial}{\partial\ep}\frac{1}{T}\right]+\\
         (mn)^2\left[\eta \frac{\partial}{\partial n}\frac{\mu}{T} +
\left.   mn\lambda \left(\frac{\partial}{\partial n}\frac{1}{T}\right)^2\right] \right)\geq0
\end{multline*}

One can see, that the first four lines of the above expression are positive, term by term due to the thermodynamic conditions (\ref{nrtdstab}) and (\ref{poscoeff}). The positivity of the last two lines can be verified if we recognize that the expression is a second order polynomial of $(\ep+p)$. The coefficient of the quadratic term is positive and the discriminant of the polynomial simplifies to the following form:
\begin{gather*}
  4\eta(mn)^2\left(-\eta+mn\lambda \frac{\partial}{\partial\ep}\frac{1}{T}\right)\Delta\leq0, \end{gather*}

As the discriminant is negative, the polynomial is positive.

Therefore we have proved the linear stability of the thermodynamic equilibrium of  Fick-Fourier-Navier-Stokes fluids. The linearized equations around the homogeneous equilibrium of these fluids are asymptotically stable.

\section{Special relativistic fluids}

\subsection{Balances of particle number, energy and momentum}

For the metric (Lorentz form) we use the $g^{ab}= diag(-1,1,1,1)$ convention and we use a
unit speed of light $c=1$, therefore for a four-velocity $u^a$ we have $u_a u^a = -1$.
$\Delta^a_{\;\;b} = g^a_{\;\;b} + u^a u_b$ denotes the $u$-orthogonal projection.
We perform the stability investigations in a general frame independently of the conventions used by Eckart or Landau and Lishitz.
 
The particle number flow can be expressed by the local rest frame
quantities as
 \be
 N^a = n u^a+j^a.
 \label{N}\ee

Here $n= -u_a N^a$ is the {\em particle density} and $j^a=\Delta^a_b N^b $ is the {\em particle flux} in a comoving frame.

The particle number conservation is described by
 \be
 \partial_a N^a = \dot{n} + n\partial_a u^a + \partial_a j^a = 0,
 \label{nbal}\ee

\noindent where $\dot{n} = \frac{d n}{d\tau} = u^a
\partial_a n$ denotes the derivative of $n$ with respect to the proper time
$\tau$.

The energy-momentum density tensor is given with the help of the
rest-frame quantities as
 \be
  T^{ab} = {e} u^a u^b + u^a q^b +
    u^b {q}^a + P^{ab},
 \label{T}\ee

\noindent where $e = u_a u_b T^{ab}$ is the {\em density of the energy}, $q^b =
-u_a\Delta^b_{\;\;\gamma} T^{a\gamma}$ is the  {\em energy flux} or {\em heat flux},
${q}^a = -u_b\Delta^a_{\;\;c}T^{c b}$ is the {\em momentum density} and $P^{ab}=
\Delta^a_c \Delta^b_d T^{cd}$ is the {\em pressure (stress) tensor}. The
momentum density, the energy flux and the pressure are spacelike in the comoving frame,
therefore $u_a q^a = 0$ and $u_a P^{ab} = u_b P^{ab} = 0^b$. Let us emphasize that this form of the energy-momentum tensor is completely general, but expressed by
the local rest frame quantities. The energy-momentum tensor is symmetric, because we
assume that the internal spin of the material is zero. In this case the heat flux and the
momentum density are equal. However, the difference in their physical meaning is a key
element of our train of thoughts. Heat is related to dissipation of energy but momentum
density is not, therefore this difference should appear in the corresponding
thermodynamic framework.

Now the conservation of energy-momentum  $\partial_b T^{ab} = 0$ is expanded to \be
 \partial_b T^{ab} =
    \dot{e} u^a + e u^a \partial_b u^b +e\dot{u}^a +
    u^a\partial_b q^b +q^b \partial_b u^a +
    \dot{{q}}^a + {q}^a \partial_b u^b +
    \partial_b P^{ab}=0.
\label{E-Ibal}\ee

Its timelike part in the local rest frame gives the balance of the
energy $e$
 \be
    -u_a\partial_b T^{ab} =
        \dot{e} + e \partial_a u^a +
        \partial_a q^a + {q}^a \dot{u}_a +
        P^{ab} \partial_b u_a = 0.
 \label{ebal}\ee

The spacelike part in the local rest frame describes the balance of
the momentum
 \be
    \Delta^a_{\;\;c}\partial_b T^{cb} =
        e\dot{u}^a  +
        {q}^a \partial_b u^b +
        q^b \partial_b u^a +
        \Delta^a_{\;\;c} \dot{{q}}^c +
        \Delta^a_c\partial_b P^{bc}
    = 0.
 \label{ibal}\ee

\section{Thermodynamics}

The entropy density and flux can also be combined into a
four-vector, using local rest frame quantities
 \be
  S^a = s u^a + J^a,
 \label{S}\ee
\noindent where $s= -u_a S^a$ is the  {\em entropy density} and $J^a = S^a - u^a s =
\Delta^a_{\;\;b} S^b$ is the {\em entropy flux}. The entropy flux is $u$-spacelike,
therefore $u_a J^a = 0$. Now the Second Law of thermodynamics is translated to the
following inequality
 \be
 \partial_a S^a = \dot{s} + s\partial_a u^a +
    \partial_a J^a \geq 0.
 \label{sbal}\ee

Relativistic thermodynamic theories assume that the entropy is a
function of the local rest frame quantities, because the
thermodynamic relations reflect general properties of local material
interactions. The most important assumption is that the entropy is a
function of the local rest frame energy density, the time-timelike
component of the energy momentum tensor according to the velocity
field of the material \cite{Eck40a3}. Definitely the
thermodynamics cannot be related to an external observer, therefore
the dependence on the relative kinetic energy is excluded. This
interpretation of $e$ in \re{T} is supported by the form of the
energy balance \re{ebal}, where the last term is analogous to the
corresponding internal energy source (dissipated power) of the
nonrelativistic theories.

In nonrelativistic fluids the internal energy is the difference of
the conserved total energy and the kinetic energy of the material.
However, also in nonrelativistic theories the constitutive relations
must be objective in the sense that they cannot depend on an
external observer, the thermodynamic framework should produce frame
independent material equations. (This apparent contradiction of
classical physics is eliminated by different sophisticated methods
and lead to such important concepts as the configurational forces
or/and virtual power \cite{Mau99b,Gur00b,Sil97b}). However, without
distinguishing the energy related to the flow of the material from
the total energy one mixes the dissipative and nondissipative
effects. The wrong separation leads to generic instabilities of the
corresponding theory.

Our candidate of the relativistic internal energy is related to the energy vector defined
by $E^a = -u_b T^{ab} = e u^a + q^a$. The energy vector embraces both the total rest
frame energy density and the rest frame momentum. Therefore its absolute value $\ep=\|E\|=
\sqrt{-E_a E^a} = \sqrt{e^2 - q_a q^a}$ seems to be a reasonable choice of the scalar
internal energy. Its series expansion, when the energy density is larger than the
momentum density is analogous to the corresponding nonrelativistic definition
\begin{equation}
  \ep =\sqrt{e^2 - q_a q^a} \approx e -
    \frac{{\bf q}^2}{2 e} + ...
\label{appr}\end{equation}

Thermodynamic calculations based on the Liu procedure support this assumption
\cite{Van08a}. Let us emphasize, that our candidate of internal energy is not related to
any external reference frame, only to the velocity field of the material. In a
Landau-Lifschitz frame the energy vector is timelike.

Assuming that the entropy density is the function of the internal energy and the particle
number density $s(e,q^a,n)=\hat{s}(\sqrt{e^2 - q_a q^a},n)$ leads to a modified form of
the thermodynamic Gibbs relation and the potential relation for the densities as follows
 \be
  de - \frac{q^a}{e} dq_a = T  ds  + \mu dn,
  \qquad \text{and} \qquad
  e - \frac{q^aq_a}{e} = T s - p + \mu n.
\label{eqt}\ee

Here $T$ is the temperature, $p$ is the pressure and $\mu$ is the
chemical potential. Equivalently the Gibbs relation gives the
derivatives of the entropy density as follows
 $$
 \left. \frac{\partial s}{\partial e}\right|_{(q^a,n)}
    = \frac{1}{T}, \qquad
 \left. \frac{\partial s}{\partial q^a}\right|_{(e,n)}
    = -\frac{q_a}{eT}, \qquad
  \left. \frac{\partial s}{\partial n}\right|_{(q^a,e)}
 = -\frac{\mu}{T}.
$$

For the entropy flux we assume the classical form
 \be
 J^a = \frac{q^a}{T}- \frac{j^a \mu}{T}.
 \label{sf}\ee

Now we substitute the energy balance \re{ebal} and the particle
number balance \re{nbal} into the entropy balance \re{sbal} and we
arrive at the following entropy production formula:
 \bea
  \partial_a S^a &=&
    \dot{s}(e,q^a,n) +
    s \partial_a u^a +
    \partial_a J^a \nonumber \\
  &=& \frac{\partial s}{\partial e}\dot{e} +
    \frac{\partial s}{\partial q^a} \dot{q}^a +
    \frac{\partial s}{\partial n}\dot{n} +
    s \partial_a u^a +
    \partial_a \frac{q^a}{T} -\partial_a \frac{j^a\mu}{T}\nonumber \\
  &=& -\frac{1}{T }(e\partial_a u^a +
        \partial_a q^a +
        q^a \dot{u}_a +
        P^{ab}\partial_b u_a) -
    \frac{q^a}{T e}\dot{q}_a +
    s\partial_a u^a  \nonumber\\
  && +\frac{\mu}{T} (n \partial_a u^a + \partial_a j^a)+
    q^a \partial_a\frac{1}{T } +
    \frac{1}{T} \partial_a q^a -
     j^a \partial_a\frac{\mu}{T} -
    \frac{\mu}{T} \partial_a j^a \nonumber\\
  &=& -j^a \partial_a\frac{\mu}{T} 
     -\frac{1}{T }\left(P^{ab} -
    \left(p+ \frac{q^aq_a}{e} \right)\Delta^{ab}\right)
        \partial_a u_b  \\
  && +{q}^a\left(\partial_a\frac{1}{T}
        -\frac{\dot{u}_a}{T} - \frac{\dot{q}_a}{e T}
       \right)\geq 0.\nonumber
 \label{cld}\eea

According to this quadratic expression and the potential relation in
\re{eqt} the {\em viscous pressure} is given by
$$
\Pi^{ab} = P^{ab}- \left(p-\frac{q^2}{e} \right)\Delta^{ab}
$$

Therefore the \re{cld} entropy production can be written as
 \be
 -j^{a}\partial_a\frac{\mu}{T}
 -\frac{1}{T}\Pi^{ab}\partial_a u_b
 +q^a\left(\partial_a\frac{1}{T}-
    \frac{\dot{u}_a}{T}
    -\frac{\dot{q}_a}{eT}\right) > 0
\label{entrpr}\ee

In isotropic continua the above entropy production results in the
following constitutive functions assuming a linear relationship
between thermodynamic fluxes and forces
 \bea
  q^a &=&
    \lambda \Delta^{ab}
    \left(\partial_b\frac{1}{T} -
         \frac{\dot{u}_b}{T} -
        \frac{\dot{q}_b}{Te} \right),\label{fo}\\
  j^a &=&-\xi \Delta^{ab}\partial_b\frac{\mu}{T}, \label{fi}\\
  \Pi^{ab} &=&
    -2\eta \langle\partial^a u^b\rangle-\eta_v \partial_b u^b\Delta^{ab}, \label{newt}
\eea
\noindent where the bracket denotes the symmetric traceless part of the spacelike tensor
$$
\langle\partial^a u^b\rangle=
        \Delta^a_c\Delta^b_d\left(\frac{\partial^c u^d+\partial^d u^c}{2}-
        \frac{1}{3} \partial_eu^e \Delta^{cd}\right).
$$

Here \re{fo}, \re{fi} and \re{newt} are the relativistic generalizations of the
Fourier law of heat conduction, the Fick law of diffusion and the Newtonian viscous pressure
function. The shear and bulk viscosity coefficients, $\eta$ and
$\eta_v$, the heat conduction coefficient $\lambda$, and the diffusion coefficient $\xi$ are non negative, according to the inequality
of the entropy production \re{fo}. We may introduce a relaxation
time $\tau = \lambda/e$ in \re{fo}, as usual in second order
theories.

The equations \re{nbal}, \re{ebal} and \re{ibal} are the evolution equations of a relativistic heat
conducting ideal fluid, together with the constitutive functions \re{fi},
\re{newt} and the
relaxation type equation \re{fo}. As special cases we can get the relativistic
Navier-Stokes equation substituting \re{newt} into \re{ibal} and assuming $q^a = 0$, or
the equations of relativistic heat conduction solving together \re{fo} and \re{ebal}
assuming that $\Pi^{ab}=0$ and $u^a = const.$.

\section{Linear stability}

In this section we investigate the linear stability of the
homogeneous equilibrium of the equations \re{nbal}, \re{ebal} and
\re{ibal} together with the constitutive relations \re{fo}-\re{newt}. Similar calculations are given by Hiscock and Lindblom both
for Eckart fluids \cite{HisLin85a} and Israel-Stewart fluids
\cite{HisLin87a}.

\subsection{Equilibrium}

The equilibrium of the above set of equations is defined by
vanishing proper time derivatives and by zero entropy production
with vanishing thermodynamic fluxes
 \be
 \Pi^{ab} = 0,  \quad j^a=0, \quad \textrm{and} \quad q^a = 0.
\label{et1}\ee

Therefore according to the balances and the constitutive functions
the equilibrium of the fluid is determined by
 \begin{gather}
 n = \text{const.} \quad e = \text{const.} \quad \Rightarrow
  \quad T= \text{const.}, \quad \mu = \text{const.}, \quad
    p= \text{const.}, \label{eq1}\\
  \partial_a u^{a} = 0, \qquad
    \partial_a u_b + \partial_b u_a =0. \label{eq2}
 \end{gather}

In addition to the above conditions we require a homogeneous
equilibrium velocity field
 \be
  u_a = \text{const.}
 \label{e3}\ee

\subsection{Linearization}

We denote the equilibrium fields by overline and the perturbed fields by $\delta$
as $Q = Q_0 +\delta Q$. Here $Q$ stands for $n$, $e$, $u^a$, $q^a$, $j^a$ and $\Pi^{ab}$. The
linearized equations \re{nbal}, \re{ebal}, \re{ibal}, \re{fo}, \re{fi} and \re{newt} around the
equilibrium given by \re{et1}-\re{e3} become
\bea
 0 &=& \dot{\delta n} +
        \ol n \partial_a \delta u^a +\partial_a\delta j^a, \label{lnbal}\\
 0 &=& \dot{\delta e} +
        (\ol e+\ol p) \partial_a \delta u^a +
        \partial_a \delta q^a, \label{lteb}\\
 0 &=& (\ol e+\ol p)\dot{\delta u^a}  +
        \Delta^{ab} \partial_b \delta p +
        \dot{\delta q^a} +
        \Delta^a_c \partial_b \delta\Pi^{cb},
        \label{ltib}\\
 0 &=& \delta q^a-
        \lambda \Delta^{ab}\left(\partial_b \delta\frac{1}{T} -
         \frac{\dot{\delta u}_b}{T} -
        \frac{\dot{\delta q}_b}{eT} \right), \label{lfum}\\
 0 &=& \delta j^a+ 
        \xi\ \Delta^{ab}\partial_b \delta\frac{\mu}{T} , \label{lfim}\\
 0 &=&  \delta\Pi^{ab} +
        {\eta}_v \partial_c \delta u^c \Delta^{ab} +
        \eta \Delta^{ac}\Delta^{bd}(
            \partial_\gamma\delta u_d +\partial_d\delta u_c-
            \frac{2}{3}\partial_e\delta u^e\Delta_{cd}).
        \label{lnewm}
\eea
The perturbation variables satisfy the following properties inherited from the linearization of the original ones
 $$
 0 = u^a \delta q_a =
 u^a \delta u_a =
 u^a \delta \Pi_{ab} =
 \delta \Pi_{ab} -  \delta \Pi_{ba}
$$

For the stability investigations we introduce exponential plane-wave
solutions of the perturbation equations: $\delta Q =  Q_0 e^{\Gamma t + i k x}$, where
$Q_0$ is constant and $t$ and $x$ are two orthogonal coordinates in Minkowski spacetime.
As our equilibrium background state is a fluid at rest we put $u^a\partial_a
=\partial_t$.

With these assumptions the set of perturbation equations follow as
 \bea
 0 &=& \Ga{\delta n} + i k n \delta u^x+ik\delta j^x,
    \label{lin1}\nonumber\\
 0 &=& \Ga{\delta e} +(e+p) i k \delta u^x +i k \delta q^x,
    \label{lin2}\nonumber\\
 0 &=& \Ga(e+p)\delta u^x + i k (\partial_e p \delta e +
    \partial_n p \delta n)+ \Gamma \delta q^x + i k \delta\Pi^{xx},
    \label{lin3x}\nonumber\\
 0 &=& \Ga(e+p)\delta u^y + \Gamma\delta q^y + i k \delta\Pi^{xy},
    \label{lin3y}\nonumber\\
 0 &=& \Ga(e+p)\delta u^z + \Gamma\delta q^z + i k \delta\Pi^{xz},
    \label{lin3z}\nonumber\\
 0 &=& \delta q^x - i k \lambda\left(\partial_e \frac{1}{T} \delta e +
    \partial_n \frac{1}{T} \delta n\right)+
    \frac{\lambda}{T} \Gamma \delta u^x +
   \frac{ \lambda }{Te} \Gamma \delta q^x,
    \label{lin4x}\nonumber\\
 0 &=& \delta q^y +
    \frac{\lambda}{T} \Gamma \delta u^y +
    \frac{\lambda }{Te} \Gamma \delta q^y,
    \label{lin4y}\nonumber\\
 0 &=& \delta q^z +
    \frac{\lambda}{T}  \Gamma \delta u^z +
    \frac{\lambda }{Te} \Gamma \delta q^z,
    \label{lin4z}\nonumber\\
  0 &=& \delta j^x + i k \xi\left(\partial_e \frac{\mu}{T} \delta e +
    \partial_n \frac{\mu}{T} \delta n\right)
    \label{lin5x}\nonumber\\
      0 &=& \delta j^y=\delta j^z,
    \label{lin5zy}\nonumber\\
 0 &=&  \delta\Pi^{xx} + ik \tilde{\eta}\delta u^x,
    \label{lin6xx}\nonumber\\
 0 &=&  \delta\Pi^{xy} + ik \eta\delta u^y,
    \label{lin6xy}\nonumber\\
 0 &=&  \delta\Pi^{xz} + ik \eta\delta u^z,
    \label{lin6xz}\nonumber\\
 0 &=&  \delta\Pi^{yy} + ik \tilde{\eta}_v\delta u^y,
    \label{lin6yy}\nonumber\\
 0 &=&  \delta\Pi^{zz} + ik \tilde{\eta}_v\delta u^z,
    \label{lin6zz}\nonumber\\
 0 &=&  \delta\Pi^{zy}.
    \label{lin6zy}
\eea

Here we have introduced shortened notations for $\tilde{\eta} =
\eta_v +\frac{4}{3}\eta$ and $\tilde{\eta}_v =
\eta_v -\frac{2}{3}\eta$ and for the partial derivatives of the thermodynamic quantities as $\partial_e = \frac{\partial}{\partial e}$ and $\partial_n = \frac{\partial}{\partial n}$. We can put the equations above into the
following matrix form
 \be
  M^A_{\;\;B} \delta Q^B = 0.
\label{meq}\ee

Here $\delta Q^B$ represents the list of fields which describe the
perturbation of the fluid:
\begin{gather*}
 \delta Q =
    (\delta n, \delta e, \delta u^x, \delta q^x, \delta \Pi^{xx},\delta j^x;\\     \delta u^y, \delta q^y, \delta \Pi^{xy},\delta \Pi^{yy};\;\;
    \delta u^z, \delta q^z, \delta \Pi^{xz},\delta \Pi^{zz};\\
    \delta j^y, \delta j^z, \delta \Pi^{yz}).
\end{gather*}

Then the 17x17 matrix {\bf M} can be written in the block diagonal
form
 \be
  {\bf M} = \begin{pmatrix}
  {\bf N} & 0       & 0       & 0       \\
  0       & {\bf R} & 0       & 0       \\
  0       & 0       & {\bf R} & 0       \\
  0       & 0       & 0       & \bf I \end{pmatrix},
 \label{M}\ee

\noindent where {\bf I} is a 3x3 identity matrix and the submatrices {\bf R} and {\bf N} are defined as
follows
 \be
  {\bf R} = \begin{pmatrix}
  (e+p)\Ga          & \Ga                       & ik    & 0\\
  \frac{\lambda}{T} \Ga      & 1+ \Ga\frac{\lambda}{Te}  & 0     & 0\\
  ik\eta            & 0                         & 1     & 0\\
  ik\tilde{\eta}_v  & 0                         & 0     & 1\\
  \end{pmatrix},
 \label{R}\ee

\be
  {\bf N} = \begin{pmatrix}
  \Ga & 0   & ikn      & 0    & 0 & ik \\
  0   & \Ga & ik(e+p)  & i k  & 0 &  0 \\
  ik\partial_n p        & ik\partial_e p  & \Ga(e+p) & \Ga  & ik & 0 \\
  -ik\lambda\partial_n \frac{1}{T} & ik\lambda\partial_e \frac{1}{T}
        &  \Ga \frac{\lambda}{T}     & 1+\frac{\lambda}{eT}\Ga & 0 & 0\\
  0   & 0   & ik\tilde{\eta}  & 0 & 1 & 0\\
 ik\xi\partial_n \frac{\mu}{T} & ik\xi\partial_e \frac{\mu}{T}
        & 0 & 0 & 0 & 1
   \end{pmatrix}.
 \label{Nm}\ee

Exponentially growing plane-wave solutions of \re{meq} emerge
whenever $\Gamma $ and $k$ satisfy the dispersion relation
 \be
 \det {\bf M} = (\det {\bf N})(\det {\bf R})^2=0
\ee with a positive real $\Gamma$. The roots of this equation are
the roots obtained by setting the determinants of either {\bf N} or
{\bf R}  to zero.

The determinant of {\bf R} gives the condition
 $$
 \lambda  \frac{p}{eT} \Ga^2 +
 \left(e+p+ k^2 \frac{\tilde{\eta}\lambda}{Te}\right)\Ga +
 \tilde{\eta} k^2 = 0.$$

The real parts of the roots of this polynomial are negative because
the coefficients of both the linear and the quadratic term are
positive.

The determinant of {\bf N} gives the following dispersion relation
 \begin{gather*}
 \Ga^4 \frac{ \lambda p}{Te}  + \\
 \Ga^3 \left(e + p + 
        k^2\frac{\lambda}{eT}\left(\tilde{\eta} +
                p\xi \partial_n\frac{\mu}{T}\right)\right) + \\
 \Ga^2k^2 \left(\tilde{\eta} +
        (e+p)\xi\partial_n\frac{\mu}{T}+ 
        \lambda n \partial_n\frac{1}{T}+
        \frac{\lambda }{eT}\left(n \partial_n p + 
                p \partial_ep+
                k^2\tilde\eta\xi \partial_n\frac{\mu}{T}\right) \right) +\\
 \Ga k^2 \left((e+p)\partial_e p + n \partial_n p  +
        k^2 \tilde{\eta}\left(\xi \partial_n\frac{\mu}{T}\!- 
        \lambda \partial_e\frac{1}{T}  \right) \!+
        k^2 \frac{p\lambda\xi}{eT}\left(\partial_ep \partial_n\frac{\mu}{T}-
                \partial_e\frac{\mu}{T}\partial_np  \right)\right)+ \\
 k^4\left(\lambda n \left(  
        \partial_e p \partial_n \frac{1}{T}-
        \partial_n p\partial_e \frac{1}{T}  \right)+
        \xi(e+p)\left(  
        \partial_e p \partial_n \frac{\mu}{T}-
        \partial_n p\partial_e \frac{\mu}{T}  \right)\right. +\\
     \left.k^2\tilde\eta\lambda\xi\left(  
         \partial e \frac{\mu}{T} \partial_n \frac{1}{T}-
        \partial_n \frac{\mu}{T} \partial_e \frac{1}{T}  \right)\right)= 0.
 \end{gather*}

According to the Routh-Hurwitz criteria \cite{KorKor00b}, the real
parts of the roots of a fourth order polynomial $a_0 x^4+a_1 x^3 +
a_2 x^2 + a_3 x + a_4=0$ are negative whenever \bea
 a_0 &>& 0, \nonumber\\
 a_1 &>& 0, \nonumber\\
 a_1 a_2 - a_0 a_3&>& 0, \nonumber \\
 (a_1 a_2 - a_0 a_3)a_2 - a_4 a_1^2 &>& 0.
\label{RH}\eea

We can see, that the first two conditions of \re{RH} are fulfilled
according to the Second Law, the nonnegativity of the entropy
production and the thermodynamic stability, the concavity of the entropy function.

Let us observe that one could define two set of intensives according to the derivatives of the entropy by the total and the internal energies \cite{BirAta08a}. However, in the stability investigations we need only the equilibrium values of the thermodynamic relations, when $q^a=0$, therefore the two set of intensives as well as their partial derivatives coincide. Hence the thermodynamic stability is required in the same form as it was in the non-relativistic case, but the state variables are the particle density $n$ and the total energy density $e$, instead of the density and the internal energy.
\re{nrtdstab}  \bea
    \partial_e \frac{1}{T} < 0, \label{s1} \\
    \partial_n \frac{\mu}{T} > 0, \label{s2} \\
    \Delta:=\partial_e \frac{1}{T}  \partial_n \frac{\mu}{T} -
        \partial_n \frac{1}{T}  \partial_e \frac{\mu}{T} \leq 0. \label{s3}
  \eea
\noindent \\ As in the nonrelativistic case the following identities
are useful  \bean
     \partial_e p &=& -(e+p)T\partial_e \frac{1}{T} + n T\partial_e \frac{\mu}{T}, \\
     \partial_n p &=& -(e+p)T\partial_n \frac{1}{T} + n T\partial_n \frac{\mu}{T},\\
     \partial_n \frac{1}{T} &=& -\partial_e \frac{\mu}{T}. 
\eean

We can see that the first two inequalities of \re{RH} are satisfied and the third inequality can be written as
\begin{multline*}
a_1a_2-a_0a_3 =\\
    k^2\left[\eta(e+p) + 
        (\lambda n^2+\xi(e+p)^2)\partial_n\frac{\mu}{T}\right]+\\
    k^4\left[\frac{\eta\lambda}{e^2T}\left(e\eta +
        \lambda\underline{\left(n^2\partial_n\frac{\mu}{T}-
                2 np \partial_n\frac{1}{T} -
                np^2\partial_e\frac{1}{T}\right)}\right)+\right.\\
         \left.\xi\lambda\left\{2\eta e(e+p)\partial_n\frac{\mu}{T} +
                \lambda ep^2\left(\partial_n\frac{1}{T}\right)^2 +
                \lambda p\left(p\partial_n\frac{1}{T}-
                        n\partial_n\frac{\mu}{T}\right)^2\right\}\right]+\\
    k^6\left[\frac{\eta\xi\lambda^2}{(eT)^2}\partial_n\frac{\mu}{T}
        \left(\eta +\xi p \partial_n\frac{\mu}{T}\right)\right]\\
\end{multline*}

This expression is positive term by term. Only the underlined part requires separate investigation. There one can recognise a second order polynomial of $n$ with the negative discriminant
$$
 4n^2 \Delta \leq 0.
$$

In Appendix B  we show that the positivity of the fourth inequality of \re{RH}  does not need any more condition beyond the concavity of the entropy and the positivity of the transport coefficients. 

Therefore we conclude that the homogeneous
equilibrium of the relativistic heat conducting, viscous relativistic
fluids is stable. We did not need to exploit any special additional
stability conditions beyond the well known thermodynamic
inequalities. This is in
strong contrast to the Eckart theory, where one encounter generic instabilities and to the M\"uller-Israel-Stewart theory, where one should
assume additional complicated conditions \cite{HisLin87a} and Appendix A.
  Our statement is valid in arbitrary flow-frame, we did not need fixed the flow neither to the Eckart nor by the Landau-Lifshitz conditions. 

\section{Conclusion}

The suggested relativistic form of the internal energy depends on
the momentum density, therefore the entropy function is the function
of the momentum density, too. Moreover, in the first
approximation we have a regular second order theory with only one
additional quadratic term in the entropy four vector according to (\ref{appr}). However, in contrast to any other extended theories there
was no need to introduce additional parameters, the coefficient of
the quadratic term, and therefore the relaxation time in the
generalized Fourier equation, is fixed. Therefore our theory of dissipative  relativistic fluids can be considered as the minimal stable extension of the theory of Eckart. 

We have investigated the analogous non-relativistic theory to emphasize the differences and similarities of the conceptual questions. However, it is well known, that the suggestion of Brenner is untenable as it is, because the conservation of the moment of momentum requires the equality of the mass current and the momentum density, similarly to the relativistic case, where the time-spacelike  and space-timelike parts of the energy momentum are equal in any frame as a consequence of the symmetry of the energy-momentum density in absence of internal  moment of momentum \cite{KosLiu98a,Liu08c}. However, as the relativistic case demonstrates clearly, the
question is not clarified completely, one can figure out several ideas to improve  and further generalize non-relativistic fluid dynamics e.g. as the nonrelativistic limit of the recent suggestion.  

There are arguments that a suitable chosen flow-frame would eliminate the instabilities of the Eckart theory \cite{KosLiu00a,TsuKun08m}. However, the simplest choice to fix the flow is the one of Eckart (coming from the concept of the non-relativistic baricentric velocity) with a vanishing particle  flux. The other natural choice, to fix the flow to the energy flux according to Landau and Lifshitz \cite{LanLif59b,Cse94b}, leads to a heat flux that is strictly connected to diffusion.  Here we have shown that from a stability point of view the flow frames are not necessary ingredients of a dissipative relativistic theory of fluids. 


\section{Acknowledgment}

This work has been supported by the Hungarian National Science Fund
OTKA (T48489).
The author is thankful to T. Matolcsi for many important comments and suggestions. 
\newpage

 \section{Appendix A: Stability conditions of the Israel-Stewart theory}

The Israel-Stewart theory is based on the following form of the entropy four vector
$$
S^a = \left( s- \frac{1}{2 T}\left(\beta_0\Pi^{2}+\beta_1q^bq_b +\beta_2 \pi^a_{\;b}\pi^b_{\;a}\right) \right)u^a+ 
        \frac{1}{T}\left(\alpha_0 \Pi q^a+\alpha_1 \pi^a_{\;b}q^b\right).
$$
Here $\Pi = \Pi^a_{\;a}$ and $\pi^a_{\;b} = \langle \Pi^a_{\;b}\rangle$ are the trace and the symmetric traceless part of the dissipative pressure and $\beta_0,\beta_1, \beta_2, \alpha_0,\alpha_1$ are the scalar valued Israel-Stewart coefficients. Stability conditions were calculated by Hiscock and Lindblom in \cite{HisLin88a}. The independent variables of the thermodynamic functions are different that we have used in the previous calculations. They use specific entropy \(\mathfrak{s}=s/n\)  and find useful to introduce \(\theta=\frac{\mu}{nT}\) as independent variable. They require the positivity of the following quantities as conditions of the stability (formulas (53)-(60), with \(\lambda=1\)):

\begin{eqnarray}
 \Omega_1 &=&
        \frac{1}{(e+p)}\left. \frac{\partial e}{\partial p} \right|_\mathfrak{s},\\  \Omega_2 &=&
        \frac{1}{(e+p)}\left.\frac{\partial e}{\partial \mathfrak{s}} \right|_p         \left.\frac{\partial p}{\partial \mathfrak{s}} \right|_\theta,\\
 \Omega_3 &=&
        (e+p)\left[ 1- \left. \frac{\partial e}{\partial p} \right|_\mathfrak{s}\right],\\
 \Omega_4 &=&
        (e+p) - \frac{2\beta_{2}+\beta_1+2\alpha_1}{2\beta_{1}\beta_{2}-\alpha_1^2},\\
 \Omega_5 &=& \beta_0,\\
 \Omega_6 &=& \beta_1-\frac{\alpha _{0}^{2}}{\beta_0}-\frac{2\alpha_1^2}{3\beta_2}-
        \frac{1}{nT^2}\left.\frac{\partial T}{\partial \mathfrak{s}} \right|_n,\\  \Omega_7 &=& \beta_1-\frac{\alpha_1^2}{2\beta_2},\\
 \Omega_8 &=& \beta_2,
\end{eqnarray}

It is worth to express the partial derivatives  of the thermodynamic quantities in the above expressions by partial derivatives with the energy density and particle number density as variables - using the previous shorthand notation - used in the our recent investigations:

\begin{eqnarray}
  \left.\frac{\partial e}{\partial p}\right|_\mathfrak{s} 
    &=& \frac{(e+p)T}{(e+p)^2\partial_eT-n^2T^2\partial_n\frac{\mu}{T}},\\
  \left.\frac{\partial e}{\partial \mathfrak{s}} \right|_p       
  \left.\frac{\partial p}{\partial \mathfrak{s}} \right|_\theta 
    &=& \frac{n^3T^3}{(e+p)^2}\partial_np\left(
    \frac{n(e+p)\partial_nT + 
    n^2T\partial_n\frac{\mu}{T}}{(e+p)^2\partial_eT-n^2T^2\partial_n\frac{\mu}{T}}\right)\\     &-&
    \frac{\mu-nT \partial_n\frac{\mu}{T}}{(e+p)\partial_nT+T\mu-nT^2\partial_n\frac{\mu}{T}}.\nonumber\\   \left.\frac{\partial T}{\partial \mathfrak{s}}\right|_n 
    &=& \frac{T}{n}\partial_eT. 
\end{eqnarray}

With the graphic representation method of inequalities of the partial derivative  of Hiscock and Lindblom \cite{HisLin83a} one can understand the physical content of the conditions. However, the meaning, the origin why one should require such a difficult set of conditions \textit{in addition} to the thermodynamic restrictions, remains obscure.

\section{Appendix B: Fourth Routh-Hurwitz condition for relativistic fluids}

The fourth  inequality in \re{RH} can be written in the following form
\begin{multline*}
(a_1a_2-a_0a_3)a_3-a_1^2a_4 =\\
k^4T\left[
        \left(\lambda n^2 + \xi(e+p)^2\right)\left(n\partial_n\frac{\mu}{T}-
                (e+p)\partial_n\frac{1}{T}\right)^2 +\right.\\
         \left.(e+p)\eta\left(\underbrace{
                n^2\partial_n\frac{\mu}{T}-
                2 n(e+p) \partial_n\frac{1}{T} -
                n(e+p)^2\partial_e\frac{1}{T}
                }\right) \right]+\\
  k^6\left[
        \frac{\lambda\eta^2}{e}\left(-e(e+p)\partial_e\frac{1}{T}\right)+
        \left(\underbrace{
                n^2\partial_n\frac{\mu}{T}-
                2 n(e+p) \partial_n\frac{1}{T} -
                n(e+p)^2\partial_e\frac{1}{T}
                }\right)+\right.\\
        \frac{\lambda\eta^2}{e}\left(\underline{
                        e^2\left(p^2\left(\partial_e\frac{1}{T}\right)^2+
                        2 n p \partial_e\frac{1}{T}\partial_n\frac{1}{T}+
                        2n^2\left(\partial_n\frac{1}{T}\right)^2\right)+
                        }\right.\\ \underline{
                   e\left(2p^3\left(\partial_e\frac{1}{T}\right)^2+
                        6 n p^2 \partial_e\frac{1}{T}\partial_n\frac{1}{T}+
                        6n^2p\left(\partial_n\frac{1}{T}\right)^2-
                        2n^3 \partial_n\frac{1}{T}\partial_n\frac{\mu}{T}\right)+
                        }\\\left. \underline{
                    \left(n^2\partial_n\frac{\mu}{T}-
                        2np\partial_n\frac{1}{T}-
                        p^2\partial_e\frac{1}{T}\right)^2
                        }\right)+\\ 
            \xi\left\{\eta^2(e+p)\partial_n\frac{\mu}{T}+\right.\\
                \frac{\eta\lambda}{e}\left(\underline{
                        n^2(3e+2p)2\left(\partial_n\frac{1}{T}\right)^2 -
                        n4(e+p)^2\partial_n\frac{1}{T}\partial_n\frac{\mu}{T}+
                        (e+p)^2(3e+p)2\left(\partial_n\frac{1}{T}\right)^2-
                        }\right.\\\left.\underline{
                        (e+p)^2 p \partial_e\frac{1}{T}\partial_n\frac{\mu}{T}
                        }\right)+\\
                 \left. \frac{\lambda^2 p}{e^2}\left(n\partial_n\frac{\mu}{T}-
                        (e+p)\partial_n\frac{1}{T}\right)^2
                        \left(\underbrace{
                        n^2\partial_n\frac{\mu}{T}-
                        2 np \partial_n\frac{1}{T} -p^2\partial_e\frac{1}{T}
                        }+
                        ep\partial_e\frac{1}{T}\right)\right\}+\\
             \left.\xi^2\left\{\eta(e+p)^2\left(\partial_n\frac{\mu}{T}\right)^2+
                        \frac{\lambda}{e}p(e+p)\partial_n\frac{\mu}{T}
                                \left((e+p)\partial_n\frac{1}{T}-
                                n\partial_n\frac{\mu}{T}\right)^2\right\}
    \right]+
\end{multline*}

\begin{multline*}
k^8\left[
        \left(\frac{\eta\lambda}{eT}\right)^2\left(
                -e\eta \partial_e\frac{1}{T}
                +\lambda\left(p \partial_e\frac{1}{T}+
                        n \partial_n\frac{1}{T}\right)^2\right)+\right.\\
           \xi\left\{\frac{\lambda\eta^3}{eT}\partial_n\frac{\mu}{T}+
                \frac{\lambda^2\eta^2}{e^2T}\left(\underline{
                        n^2 2 \left(\partial_n\frac{\mu}{T}\right)^2-
                        2 n (e+2p) \partial_n\frac{1}{T}\partial_n\frac{\mu}{T}+
                        }\right.\right.\\\left.\underline{
                        3e(e+p)\left(\partial_n\frac{1}{T}\right)^2-
                        p(e+2p) \partial_e\frac{1}{T}\partial_n\frac{\mu}{T}
                        }\right)+\\
                \frac{\eta\lambda^3p}{e^3T}\left(\underline{
                        n^2\partial_n\frac{\mu}{T}\left(
                                (e-p)\left(\partial_n\frac{1}{T}\right)^2-
                                p \partial_e\frac{1}{T}\partial_n\frac{\mu}{T}\right)+
                                }\right.\\\underline{
                                2np\partial_n\frac{1}{T}\left(
                                (e+p)\left(\partial_n\frac{1}{T}\right)^2+(e+2
                                p) \partial_e\frac{1}{T}\partial_n\frac{\mu}{T}\right)+
                                }\\\left.\left.\underline{
                                 p(e+p)\partial_e\frac{1}{T}\left(
                                (-e+p)\left(\partial_n\frac{1}{T}\right)^2+
                                p \partial_e\frac{1}{T}\partial_n\frac{\mu}{T}\right)
                                }\right)\right\}\\
           \xi^2\left\{
                \frac{\eta^2\lambda}{eT}2(e+p)\left(\partial_n\frac{\mu}{T}\right)^2+\right.\\
                 \frac{\eta\lambda^2}{e^2T}p\partial_n\frac{\mu}{T}\left(\underline{
                        2n^2 \left(\partial_n\frac{\mu}{T}\right)^{2}+
                        2 n (e+2p) \partial_n\frac{1}{T}\partial_n\frac{\mu}{T}-                         (2e+p)(e+p)\left(\partial_n\frac{1}{T}\right)^2+
                        }\right.\\\left.\underline{
                        p( e+p) \partial_e\frac{1}{T}\partial_n\frac{\mu}{T}
                        }\right)-\\
                 \left.\frac{\lambda^3}{e^3T}p^3\left(
                        n\partial_n\frac{\mu}{T}-(e+p)\partial_n\frac{1}{T}\right)^2
                        \left(\partial_e\frac{1}{T}\partial_n\frac{\mu}{T}+
                                \left(\partial_n\frac{1}{T}\right)^2\right)
                    \right\}+\\
            \left.\xi^3\frac{\eta\lambda}{eT}p(e+p)\left(\partial_n\frac{\mu}{T}\right)^3
          \right]+\\
  k^{10}\left[
        \xi\frac{\eta^3\lambda^3}{e^2T^2}\left(\partial_n\frac{1}{T}\right)^2+\right.\\
        \xi^2\frac{\eta^2\lambda^2}{e^3T^2}\partial_n\frac{\mu}{T}\left\{                 e\eta\partial_n\frac{\mu}{T}+
                \lambda p\left((e-p)\left(\partial_n\frac{1}{T}\right)^2-
                        p \partial_e\frac{1}{T}\partial_n\frac{\mu}{T}
                        \right)\right\}+\\
      \left. \xi^3\frac{\eta\lambda^{2}}{e^{3}T^2}p
                \left(\partial_n\frac{\mu}{T}\right)^2
                \left\{e\eta\partial_n\frac{\mu}{T}-
                \lambda p^2\left(\partial_e\frac{1}{T}\partial_n\frac{\mu}{T}+
                        \left(\partial_n\frac{1}{T}\right)^2
                        \right)\right\}
  \right] \geq0.   
\end{multline*}

Here the first two underbraced expressions are identical second order polynomials of $e+p$ with a discriminant
$$
D_1=4 n^2 \Delta\leq 0. 
$$

Therefore the first part of the expression, the multiplier of $k^2$ is positive. Then the underlined expression is a second order polynomial of $e$, with a discriminant
$$
D_2= 4 n^2 \left(n\partial_n\frac{1}{T}+ p\partial_e\frac{1}{T} \right)^2\\         \left(-\left(n\partial_n\frac{\mu}{T}- p\partial_n\frac{\mu}{T}\right)^2+         2p^2\Delta\right)\leq 0.
$$

The second underlined and the third underbraced expressions are second order polynomials of $n$ with negative discriminants
$$
D_3 = 4(e+p)^2 \left(\partial_n\frac{\mu}{T}\right)^2 
        \left(p(3e+2p)\Delta- 
        e(5e+4p)\left(\partial_n\frac{1}{T}\right)^2\right)\leq 0,
$$
and
$$
D_4= 4 p^2 \Delta\leq 0.
$$
Therefore the coefficient of $k^6$ is positive, because the other terms and coefficients in that part of the expression are nonnegative. 

Similarly, in the coefficient of $k^8$ the three underlined expressions are second order polynomials of $n$ with the following discriminants
\begin{eqnarray*}
D_5 &=& 
        4 \left(\partial_n\frac{\mu}{T}\right)^2 
        \left(2p(e+2p)\Delta- 
        e(5e+4p)\left(\partial_n\frac{1}{T}\right)^2\right)\leq 0,\\
D_6 &=& 
        4 p\left(p\Delta+ e\left(\partial_n\frac{\mu}{T}\right)^2\right)^2         \left(p\Delta+ 
              e\partial_e\frac{1}{T}\partial_n\frac{\mu}{T}\right)\leq 0,\\
D_7 &=& 
        4 \left(\partial_n\frac{\mu}{T}\right)^2 
        \left(2p(e+p)\Delta- 
        e(3e+4p)\left(\partial_n\frac{1}{T}\right)^2\right)\leq 0,      
\end{eqnarray*}
respectively. As every other term and coefficient is nonnegative in the expression, including the whole coefficient of $k^{10}$, we conclude that the whole expression is nonnegative, as we have already indicated at the end of the formula.

\bibliographystyle{unsrt}

\end{document}